
\input phyzzx

\noindent
\bf A TOPOLOGICAL STUDY OF

\noindent
INDUCED REPRESENTATION\footnote*{\rm To appear in the Proceedings of
Symmetries in Science VIII, B.Gruber(ed.) Plenum.}

\vskip 14truemm
\noindent
\hskip 26truemm
\rm Kazuhiko Odaka\footnote{\dag}{e-mail address: odaka@cc.nda.ac.jp}

\vskip 5truemm
\noindent
\hskip 26truemm
Department of Mathematics and Physics

\noindent
\hskip 26truemm
National Defence Academy

\noindent
\hskip 26truemm
Yokosuka 239 JAPAN

\vskip 10truemm
\noindent
\bf INTRODUCTION

\vskip 5truemm
\rm In 1939 E. Wigner$^{1,2}$ proposed the induced representation technique
when he was obtaining the unitary representation of the Poincar\'e group.
This technique is very useful for elementary particle physics. All elementary
particles can be corresponded to the induced representations of the
Poincar\'e group.
We have not found any exception.

The Poincar\'e group $P$ is composed of two parts:
$$P~=~T \otimes L.\eqno(1.1)$$
One part $T$ is the parallel translation in space-time and another
part $L$ is the Lorentz transformation.
The symbol $\otimes$ means the semi-direct product. We take $\hat k_\mu$
to be the generators of the translation, that is,   momentum and energy.
The summation of their square
$$\hat k_0^2 - \sum_{i=1}^3 \hat k_i^2 = \hat \kappa^2\eqno(1.2)$$
is invariant under the translation and the Lorentz transformation. $\hat
\kappa^2$ is the Casimir operator of the Poincar\'e group and so the
representations are classified by the eigenvalues of $\hat \kappa$ whose
physical meaning is the mass of the particle. These classes are as follows:
$$[M_{\pm}],~~[0_{\pm}],~~[L],~~[T]$$
where $\pm$ is sgn($k_0$) and + and $-$ indicate the particle and the
anti-particle, respectively.
The first classes correspond to the usal massive particle and the second ones
are the massless particle. We should notice that these classes do not
include the zero energy mode, which belongs to the third class $[L]$.
The last case is the tachyon.

A vector space is introduced to provide the representation in the concrete.
We take the momentum diagonal state $\mid k,\xi >$
as a vector for the representation.
 Here, $\xi$ means the spin degree. The inner product
$<k,\xi\mid k',\xi'>$ must be invariant under the Lorentz transformation.

For the massive case, the Lorentz invariant inner product is given by
$$<k,\xi\mid k',\xi'> = \delta_{\xi \xi'}{\omega_k} \delta^3
(\vec{k} - \vec{k}')\eqno(1.3)$$
where $\omega_k = \sqrt{\vec k^2 + m^2}$. The state vectors are well-defined
everywhere on the momentum space, which is equivalent to the 3-dimensional
Euclidean space. This space is topologically trivial.

For the massless case, the Lorentz invariant inner product is
$$<k,\xi\mid k',\xi'> = \delta_{\xi \xi'} k_0 \delta^3
(\vec{k} - \vec{k}')\eqno(1.4)$$
where $k_0 = \mid \vec{k} \mid$. On the occasion of defining the states, one
point is subtracted from the momentum space. Therefore, the state vectors are
on the $R^3 -\{0\}$ space, which is homotopically equivalent to $S^2$. It is
a topologically non-trivial space.

There exist many problems in the quantum field theory of the massless
particle$^3$, for example, gauge anomaly$^4$. Most problems are related with
the zero energy mode, which is subtracted from the region of definition of
the state vectors. In order to settle these problems, we must study the
induced  representaion technique on the topologically non-trivial space and
its application to quantum theories.

In 1968, furthermore, E. Mackey$^{5,6}$ generalized Wigner's technique
to other groups and he study quantm mechanics on a homogeneous space ($G/H$)
by using the technique. In his research he discoursed that there exist many
inequivalent quantizations for quntum mechanics on a topologically non-trivial
configulation space and that they can be classified according to the
representaions of $H$. This situation does not appear in the usual
approach$^7$.
 We introduce a position operator on the homogeneous space and the diagonal
state vector of this operator is taken to be one for the representation. But,
such vectors is not always defined over the homogeneous space, since it is
not guarenteed that vecor fields do not vanish anywhere.  We need to study
quantum mechanics of the case that the state vector can not be defined over
all.

In this note, we will study the problem how Wiger's argument about the
induced representaion technique is changed for the topologically non-trivial
case. Our talk was in the following order; $\S$ 2. Wigner's argument and
toplogy, $\S$ 3. Euclidean group and gauge structrue and $\S$ 4. Quantum
mechanics on a sphere.

\vskip 10truemm
\noindent
\bf WIGNER'S ARGUMENT AND TOPOLOGY

\vskip 5truemm
\rm Let us review Wigner's argument$^{1,2}$ about the induced representation
technique. The group, with which we are concerned, is
$$M~=~V \otimes G\eqno(2.1)$$
where $V$ and $G$ are an abelian group and a transformation group,
respectively.
The group law on $M=V \otimes G$ is
$$(a_2, \Lambda_2) \cdot (a_1, \Lambda_1) =
(a_2 + \Lambda_2*a_1, \Lambda_2\Lambda_1). \eqno(2.2)$$
The action of $G$ on $a$ is remained on a homogeous space ($G/H \equiv X$)
such as
$$\Lambda*a \in X.\eqno(2.3)$$
In order to obtain the representation of $G$ on the space $X$, we take the
following state of the position operator $\hat x$ on $X$,
$$\hat x\mid x,\xi > = x\mid x,\xi >.\eqno(2.4)$$
These state vectors should be single-valued. If they are degenerate, a new
degree $\xi$ is introduced to remove their degeneracy.
We consider the representaion matrix $G(\Lambda)$ on the state
vector $\mid x,\xi >$. The representaion matrix $G(\Lambda)$ is divided into
two parts as
$$G(\Lambda)~=~Q(\Lambda,\hat x)P(\Lambda)\eqno(2.5)$$
where $P(\Lambda)$ and $Q(\Lambda,\hat x)$ are diagonal for $\xi$ and $\hat x$
, respectively. They act on the state vector in the following manner,
$$P(\Lambda)\mid x,\xi >~=~\mid \Lambda*x,\xi >,~~~Q(\Lambda,\hat x)\mid x,
\xi'>~=~\mid x,\xi >Q^{\xi \xi'}(\Lambda,
x).\eqno(2.6)$$
Their product rules
$$P(\Lambda')P(\Lambda) =
P(\Lambda' \Lambda),~~~~~
Q(\Lambda',\hat x)Q(\Lambda,\Lambda'^{-1}\hat x)~=~Q(\Lambda'\Lambda,\hat x)
\eqno(2.7)$$
are derived from $G(\Lambda')G(\Lambda)=G(\Lambda'\Lambda), $ eq.(2.5)
and eq.(2.6).
$P(\Lambda)$ satisfies the properties of the representation matrix but
$Q(\Lambda,\hat x)$ does not do so.

If $\Lambda'^{-1}\hat x=\hat x,$   $Q(\Lambda',\hat x)Q(\Lambda,\hat x)=
Q(\Lambda'\Lambda,\hat x)$ and then $Q(\Lambda',\hat x)$ is the representaion
matrix. Then, Wigner define a little group by
$$\lambda l~=~l,\eqno(2.8)$$
where $l$ is some fixed point on $X$. It is obvious that the little is the
subgroup $H$. Next, he show that any element of $G$ can be reduced to the
representation of the little group and the translation on $X$  by taking a
suitable unitary transformation
$$G(\Lambda)~\simeq~U(\hat x)G(\Lambda)U^{-1}(\hat x)~=~
U(\hat x)Q(\Lambda,\hat x)
U^{-1}(\Lambda^{-1}\hat x)P(\Lambda),\eqno(2.9)$$
where $\simeq$ means the unitary equivalence. This unitary matrix $U(\hat x)$
is determined through the following procedure. The boost transformation is
introduced such as
$$\alpha_x l~=~\hat x~~~~(\alpha_x\in G).\eqno(2.10)$$
The element $\lambda_x$, given by $\alpha_x^{-1}\Lambda\alpha_{\Lambda^{-1}x}$
, satisfies $\lambda_x l~=~l$  and then any element $\Lambda$ is expressed by
$$\Lambda~=~\alpha_x \lambda_x \alpha^{-1}_{\Lambda^{-1}x}.\eqno(2.11)$$
Substituting this form into $Q(\Lambda,\hat x)$, we have
$$Q(\Lambda,\hat x)~=~Q(\alpha_x \lambda_x \alpha^{-1}_{\Lambda^{-1}x},\hat x)
{}~=~Q(\alpha_x,\hat x) Q(\lambda_x,l) Q(\alpha^{-1}_{\Lambda^{-1}x},l).
\eqno(2.12)$$
Let the unitary matrix $U(\hat x)$ in eq.(2.9) be $Q(\alpha_x,\hat x)$. It is
easy to see that
$$G(\Lambda)~\simeq~Q(\lambda_x,l)P(\Lambda).\eqno(2.13)$$
Here, $Q(\lambda_x,l)$ is just a representation matrix of the subgroup $H$
and it is called Wigner rotation. We arrive at the unitary equivalent
representation which is divided into the translation and some representaion
of the subgroup $H$.

The boost operator $\alpha_x$ moves a vector at $l$ to one at $\hat x$, and
then a vector is assigned smoothly to each point of $X$ by $\alpha_x$. It is
called a vector field over $X$. However, if a d-dimensional manifold $M$ is
topologically non-trivial, it is not necessarily possible to define d vector
fields which are linearly independent everywhere. When the d-dimensional
homogeous space $X$ does not admit d linearly independent vector fields
over $X$, the boost operator $\alpha_x$ is not well defined over $X$ and
then we can not determine the unitary matrix $U(\hat x)$. Moreover, the
invariant inner product of the $\hat x$-diagonal states includes d
$\delta$-functions, but these $\delta$-functions are not defined at
a point where d vector fields are not linearly independent. Therefore
the state vecors are not defined over all.

Wigner's technique, reviewed in this section, must be amended for
the manifold which is not parallelisable. In the next section, we take
$S^d$ as a homogeous space and show how Wigner's argument is modified.

\vskip 10truemm
\noindent
\bf EUCLIDEAN GROUP AND GAUGE STRUCTURE

\vskip 5truemm
\rm We take the Euclidean group which is composed of the translation and
the rotation group $SO(d+1)$ in the d+1 dimensional Euclidean space. When
the induced representation is obtained, we do not use the momentum operator
but the position operator $\hat x$ which is a vector in the d+1 dimensinal
Euclidean space. The summation of the square of its components $\hat x_i$;
$r^2=\sum_i^{d+1}\hat x_i^2$ is invariant under the rotation. $r$ is fixed
to be 1. The boost transformation is defined by
$$\alpha_x l =\hat x,~~~~\alpha_x \in SO(d+1),\eqno(3.1)$$
where $l$ indicates some fixed point on $S^d$.

This operator also translates the vector fields at $l$ to ones at $\hat x$.
However, it is well known that the sphere $S^d$, except for $d=1, 3$ and $7$,
does not admit d vector fields which are linearly independent everywhere on
$S^d$. Therefore, the boost operators are not well defined over all. Moreover,
the state vectors, which are the eigenstates of the opsition operator, are
not well defined over $S^d$, since the definition of their inner product
includes the $\delta$-functions which can not defined at the point where
a vector field vanishes.

Wigner's argument should be madified as follows.  Firstly, we take two charts
on $S^d$,  which are named $N$ and $S$, respectively. We introduce two fixed
points ($l^N$ and $l^S$), the boost operators ($\alpha_x^N$ and $\alpha_x^S$)
and the state vectors. The induced representation technique, shown in $\S$2,
is applied to the group $SO(d+1)$ on each chrat.
The unitary matrix $U(\hat x)$ in eq(2.9) is given by $Q(\alpha_x^N,\hat x)$
and $Q(\alpha_x^S,\hat x)$ on each chart.

Secondly, these representations on each chart must be unitary-equivalently
connected  at the same point in the overlap region such that
$$Q(\alpha_x^N,\hat x)G(\Lambda)Q(\alpha_x^N,\hat x)^{-1}
\simeq Q(\alpha_x^S,\hat x)G(\Lambda)Q(\alpha_x^S,\hat x)^{-1}.\eqno(3.2)$$
{}From the uniraty equivalence $Q(\alpha_x^N,\hat x)$ and
$Q(\alpha_x^S,\hat x)$ are required to satisfy the condition that $R(\hat x)$,
defined by $Q(\alpha_x^S,\hat x)^{-1}Q(\alpha_x^N,\hat x)$, must be
a single-valued unitary matrix which is assigned smoothly to each point
of the overlap region. $R(\hat x)$ is rewritten by $Q(\alpha_x^{S-1}\alpha_x^N
,l^S)$ and it is obvious that it is not a representation matrix. Then it is
difficult to confirm directly that the above condition is satisfied by
$R(\hat x)$.

We comment on the gauge structue in the induced representation
technique$^8$. The relation of the unitary equivalence at some point is
$$R(\hat x)G^N(\Lambda,\hat x)R(\hat x)^{-1}~=~G^S(\Lambda,\hat x),
\eqno(3.3)$$
where $G^i(\Lambda,\hat x)$ ($i=S,N$) is the representation matrix divided
into the translation and the representation of $SO(d)$. Let us consider the
infinitesimal rotation ($\Lambda=1+\omega$) and the displacement $\delta x$
on $S^d$ which is given by $x+\delta x=(1+\omega)x$. From eq.(3.3), the
unitary equivalence of the infinitesimal rotation is expressed by
$$\lim_{\delta x_\mu \rightarrow 0}{R(\hat x)G^N(1+\omega,\hat x)
R(\hat x)^{-1}-1\over \delta x_\mu}~=~\lim_{\delta x_\mu \rightarrow 0}
{G^S(1+\omega,\hat x)-1\over \delta x_\mu}\eqno(3.4).$$
Noting that
$${1\over \delta x_\mu} = {1\over \omega_{\mu\nu}x_\nu} = {x_\nu\over x^2}
{1\over\omega_{\mu\nu}},\eqno(3.5)$$
we have
$$R(\hat x)\partial_\mu R(\hat x)^{-1} + R(\hat x)A_\mu^N(\hat x)
R(\hat x)^{-1} = A_\mu^S(\hat x),\eqno(3.6)$$
where
$$R(\hat x)\partial_\mu R(\hat x)^{-1} \equiv {x_\nu\over x^2}
\lim_{\omega_{\mu\nu}\rightarrow
 0}{R(\hat x)R(\hat x+\omega \hat x)^{-1}-1\over\omega_{\mu\nu}},\eqno(3.7a)$$
$$A_\mu^i(\hat x) \equiv {x_\nu\over x^2}\lim_{\omega_{\mu\nu}\rightarrow 0}
{Q^i(1+\omega,\hat x)-1\over \omega_{\mu\nu}} ~~~~~~(i=N,S).\eqno(3.7b)$$
$\partial_\mu$ in eq.(3.7a) means the differential on the sphere.
$A_\mu^i(\hat x)$ can be regarded as gauge potentials on each chart,
since eq.(3.6) expresses the gauge transformation.

Now, $R(\hat x)$ is classified by the homotopy group $\pi_{d-1}(U(n))$,
since $R(\hat x)$ is a map $R:S^{d-1}\rightarrow  U(n)$ where $U(n)$ is a
unitary group. Then the winding (wrapping) number can be defined and if
$R(\hat x)$ satisfy the above condition, the number must be a integer.
Let us examine this point for $S^2$ and $S^4$ whose homotopy groups are
$\pi_1(U(1))=Z$ and $\pi_3(U(n))=Z$, respectively. The results for the
general cases are published elsewhere$^9$. The winding (wrapping) numbers
are given by
$$k = {1\over 2\pi}\int_{S^1}R^{-1}dR,\eqno(3.8a)$$
$$k = {1\over 24\pi^2}\int_{S^3}tr\big(R^{-1}dR\wedge R^{-1}dR\wedge R^{-1}
dR\big).\eqno(3.8b)$$
Fortunately, we can calculate these numbers without the explicit form of
$R(\hat x)$ such that
$$1+R(\hat x)^{-1}dR(\hat x) = Q(\alpha_x^{S-1}\alpha_x^N,l^S)^{-1}Q
(\alpha_{x+dx}^{S-1}\alpha_{x+dx}^N,l^S) $$
$$=Q(\alpha_x^{N-1}\alpha_x^S\alpha_{x+dx}^{S-1}\alpha_{x+dx}^N,
\alpha_x^{N-1}\alpha_x^Sl^S).\eqno(3.9)$$
Defining $\chi$ by $\alpha_x^S\alpha_{x+dx}^{S-1}$, we have
$$1+R(\hat x)^{-1}dR(\hat x)  = Q(\alpha_x^{-1}\chi\alpha_{\chi^{-1}x},l^N).
\eqno(3.10)$$
Then, we can show that $1+R(\hat x)^{-1}dR(\hat x)$ is just the Wigner
rotation.

We use the stereographic projection:
$$(S)~~~(\hat x_i,\hat x_{d+1}) = {1\over 1+\hat y^2}(2\hat y_i,\hat y^2-1),
\eqno(3.11a)$$
$$(N)~~~(\hat x_i,\hat x_{d+1}) = {1\over 1+\hat z^2}(2\hat z_i,1-\hat z^2),
\eqno(3.11b)$$
where $\hat y={\hat z\over\hat z^2}$ in the overlap region, and the winding
(wrapping) number is written dwon explicitly through the following
calculation. Firstly, taking the spinor representation of the group
$SO(d+1)$, we express the boost trsnsformation as
$$\alpha^S_x = {1\over\sqrt{1+\hat y^2}}(1+\gamma_{d+1}\sum_{i=1}^d
\gamma_i \hat y_i),\eqno(3.12a)$$
$$\alpha^N_x = {1\over\sqrt{1+\hat z^2}}(1-\gamma_{d+1}\sum_{i=1}^d \gamma_i
\hat z_i),\eqno(3.12b)$$
where $\gamma_j$ is the Hermitian matrix satisfying the Clifford algebra
of order $d+1$.
$\chi$ is given by
$$\chi = 1+i\sum_{i=1}^{d+1}{d\hat y_i\over 1+\hat y^2}\sigma_{i d+1},
\eqno(3.13)$$
where $\sigma_{ij}={1\over2i}[\gamma_i,\gamma_j]$. Then, after somewhat
lengthy calculation we get
$$\alpha_x^{N-1}\chi\alpha_{\chi^{-1}x} = 1-i\sum_{j,i=1}^d{\hat z_id\hat z_j
\over{\hat z^2+1}}\sigma_{ji}.\eqno(3.14)$$
Next, replacing ${\sigma_{ij}\over 2}$ with the generator ($S_{ij}$) of some
representation of the group $SO(d)$, we arrive at
$$Q(\alpha_x^{N-1}\alpha^S_x\alpha^{S-1}_{x+dx}\alpha_{x+dx}^N,l^N) = 1-i
\sum_{j,i=1}^d{\hat z_id\hat z_j\over{\hat z^2+1}}2S_{ji}.\eqno(3.15)$$
Taking account of the metric, we express the winding (wrapping) numbers
on the unit sphere such as
$$k = {1\over \pi}\int_{S^1}\hat z_id\hat z_jS_{ij}, \eqno(3.16a)$$
$$k = {1\over 3\pi^2}\int_{S^3} tr \Big(\hat z_{i_1}d\hat z_{j_1}
S_{i_1j_1}\wedge\hat z_{i_2}d\hat z_{j_2}S_{i_2j_2}\wedge\hat z_{i_3}
d\hat z_{j_3}S_{i_3j_3}\Big). \eqno(3.16b)$$

By the way, it is noteworthy that the winding (wrapping) can be  written
in terms of the gauge fields defined in eq.(3.7b) such that
$$k = -{1\over 4\pi}\int_{S^2}F,\eqno(3.17a)$$
$$k = -{1\over 8\pi^2}\int_{S^4}tr\Big(F\wedge F\Big),\eqno(3.17b)$$
where $F$ is the field tensor of $A$.

\vskip 10truemm
\noindent
\bf QUNATUM MECHANICS ON A SPHERE

\vskip 5truemm
\rm We apply the induced representation technique to quantum mechanics on
a d-dimensional sphere $(\simeq SO(d+1)/SO(d))^{6,8}$ and formulate its
path integral for the transition amplitude by using the semi-classical
approximation$^{10}$.

A state vector $\mid \psi(t)>$ satisfies the Schr\"odinger equation
$$i{d \over dt}\mid \psi(t)> = \hat H \mid \psi(t)>\eqno(4.1)$$
whose formal solution is $\mid \psi(t)>=e^{-i\hat Ht}\mid \psi(0)>$.
The quadratic Casimir operator of $SO(d+1)$ is taken as the Hamiltonian
$\hat H$, because it is invariant under the translation on $X$ and is simple.

We take two charts on $X$ and the following program is executed on each chart.
The coordinate operator $\hat x$ is diagonalised and the state vectors are
spanned by
$$\mid x,\xi> = G(\Lambda)\mid l,\zeta>\eqno(4.2)$$
where $G(\Lambda)$ is the representation matrix of $SO(d+1)$, $l$ is some
fixed point and $x=\Lambda*l$.

The transition amplitude $\Im (T,0)$ for a particle to start at $x_0$ at
$t=0$ and end up at $x_f$ at $t=T$, is given by
$$\Im (T,0)~=~<x_f, \xi_f\mid e^{-i\hat HT} \mid x_0,\xi_0>.\eqno(4.3)$$
We divide it as
$$<x_f, \xi_f\mid \prod_{n=1}^N e^{-i\hat Hn\tau} \mid x_0,\xi_0>,\eqno(4.4)$$
where $\tau=T/N.$ Inserting the identity
$$\sum_{\xi_n}\int d\mu(x_n)\mid x_n,\xi_n><x_n,\xi_n\mid~=~1\eqno(4.5)$$
where $d\mu(x)$ is the measure satisfying $d\mu(x)=d\mu(\Lambda*x)$, we have
$$\Im (T,0)~=~\sum_{\xi_1}\int d\mu(x_1)\cdots\sum_{\xi_{N-1}}\int
d\mu(x_{N-1})<x_f, \xi_f\mid e^{-i\hat H\tau}\mid x_{N-1},\xi_{N-1}>$$
$$\cdots<x_1, \xi_1\mid e^{-i\hat H\tau}\mid x_0,\xi_0>.\eqno(4.6)$$
Each transition amplitude
$$K(x_{n+1},\xi_{n+1};x_n,\xi_n)~=~<x_{n+1}, \xi_{n+1}\mid e^{-i\hat H\tau}
\mid x_n,\xi_n>\eqno(4.7)$$
at infinitesimal interval $\tau$ is estimated
by using the semi-classical approximation.
Noting that $\mid x_n,\xi_n> = G(\Lambda_n)\mid l,\zeta>$ where $\Lambda_n*l=
x_n$, we find
$$ K(x_{n+1},\xi_{n+1};x_n,\xi_n)~=~<x_n,\zeta_{n+1}\mid G^{-1}(\omega)
e^{-i\hat H\tau}\mid x_n,\xi_n>.\eqno(4.8)$$
where $\omega\equiv \Lambda_{n+1}\Lambda_n^{-1}$. Since the Hamiltonian
is invariant under the action of $G$, this is rewritten by
$$ K(x_{n+1},\xi_{n+1};x_n,\xi_n)~=~<x_n,\xi_{n+1}\mid e^{-i\hat H\tau}
G^{-1}(\omega)\mid x_n, \zeta_n>\eqno(4.9)$$
According to the argument in $\S$2, we have
$$K(x_{n+1},\xi_{n+1};x_n,\xi_n)~=~<x_n,\xi_{n+1}\mid e^{-i\hat H\tau}
P(\omega)^{-1}\mid x_n, \zeta_n>Q^{\xi_{n} \zeta_{n}}(\omega,x_n)^{-1}.
\eqno(4.10)$$
All possible representations of the group $SO(d+1)$, which are laveled by
$S$, are put between $<x_n,\zeta_{n+1}\mid$ and $e^{-i\hat H\tau}
P(\omega)^{-1}$. The amplitude is now given by
$$ K(x_{n+1},\xi_{n+1};x_n,\xi_n)~=~Q^{\xi_{n} \zeta_{n}}(\omega,x_n)^{-1}
\sum_S e^{-iH(S)\tau} D_{\xi_{n+1}\zeta_n}^S(\omega),\eqno(4.11)$$
where
$$D_{\xi_{n+1}\zeta_n}^S(\omega)~\equiv~<x_n,\xi_{n+1}\mid S><S\mid
P(\omega)^{-1}\mid x_n, \zeta_n>.\eqno(4.12)$$
Since the Hamiltonian is quadratic Casimir operator of the group $SO(d+1)$,
it is easy to calculate its value for the representations. $D_{\xi_{n+1}
\zeta_n}^S(\omega)$ can be expressed by the Gegenbauer polynomial and
$Q^{\xi_{n} \zeta_{n}}(\omega,x_n)$ can reduced to the representaion
matrix of the little group $H$  by applying Wigner's technique. To carry
out the summation of $S$ in eq.(4.11) corresponds to integrating of momentum
in the usual path integral formalism based on the canonical quantization.
It is very hard to carry out the summation without any approximation.

Here, we take quantum mechanics on $S^2$ as the simplest example and show
its results.  The investigation on the other cases are in progress. Now,
$G$ and $H$ are $SO(3)$ and $SO(2) (\simeq U(1))$, respectively. The
Hamiltonian $\hat H$ is given by ${1\over 2} \vec L^2$ where $\vec L$
are the generators of $SO(3)$ and their reprentations are specified by
the eigenvalues $(j,m)$ of $\vec L^2$ and $L_z$. Then the transition
amplitude is
$$K(x_{n+1};x_n,s)~=~Q^s(\omega,x_n)^{-1} \sum_{j=|s|}^{\infty}
{2l+1\over4\pi}e^{(-i{\tau j(j+1)\over 2})}d^l_{ss}(\omega),\eqno(4.13)$$
where $s$ is an integer and $d^l_{ss}(\omega)$ is a well-known representation
function.  $\omega$ is specified by the inner product $x_{n+1}\cdot x_{n}=cos
\Theta$ and we use the semi-classical approximation; $\tau \ll 1,~\omega^2
\simeq O(\tau)$.  Then after somewhat lengthy calculations we are let to
$$K(x_{n+1};x_n,s)~\simeq~{1\over 2\pi i\tau} Q^s(\omega,x_n)^{-1}
e^{({4i\over\tau}(1-cos{\Theta\over2}))} e^{(-i\tau({s^2\over2}-{1\over4}))}.
\eqno(4.14)$$

$Q^s(\omega,x_n)$, by using the semi-classical approximationis, is written
in terms of the gauge field discussed in $\S$ 3 such that
$$Q^n(\omega,x_n)~\simeq~1 + iA, \eqno(4.15)$$
where
$$A~=~{-sdx_1x_2\over1+x_3} + {sdx_2x_1\over1+x_3}. \eqno(4.16)$$
For general cases, if we use the semi-classical approximation,
$Q^{\xi_{n} \zeta_{n}}(\omega,x_n)$ can be also written in terms
of the gauge fields. For example, the instanton gauge field appears
in $S^4$. This is pointed  out from the different approach$^{11}$.
However, it is emphasized that the appearance of gauege fields is
not exact but based on the semi-classical approximation.

When we take the limit $(N\rightarrow\infty)$, the transition amplitude
(4.6) is given symbolically by
$$\Im (T,0)~=~\int D\mu(x(t))~e^{i(S_{eff}+S_{top})},\eqno(4.17)$$
where
$$S_{eff}~=~\int^T_0 dt \Bigl({1\over2}\sum_{i=1}^3({dx_i\over dt})^2 +
{1\over 4} - {s^2\over 2} \Bigr),\eqno(4.18a)$$
$$S_{top}~=~s\int^T_0 dt \Bigl({-dx_1\over dt}{x_2\over1+x_3} +
{dx_2\over dt}{x_1\over1+x_3}\Bigr).\eqno(4.18b)$$
$S_{eff}$ is the classical action of free particle on $S^2$ except for
$({1\over 4}-{s^2\over 2})$.

We can also put the similar calculation in practice on another chart.
The different result appears only in $S_{top}$ such that
$$S_{top}~=~s\int^T_0 dt \Bigl({dx_1\over dt}{x_2\over 1-x_3} - {dx_2\over dt}
{x_1\over 1-x_3}\Bigr).\eqno(4.19)$$

Quantum mechanics on $S^2$ is equal to the quantum dynamics in the background
field of s magnetic monopoles within the semi-classical approximation.

\vskip 10truemm

\noindent
\bf ACKNOWLEDGEMENTS

\vskip 5truemm
\rm We would like to thank W. Thirring for his warm hospitality
at Wiener University. We gratefully acknowledge numerous conversations
with Y. Ohnuki and S. Kamefuchi.

\vskip 10truemm
\noindent
\bf REFERENCES

\vskip 5truemm
\noindent
\rm
1.~~ E. Wigner, Ann. Math. 40(1939)149.

\noindent
2.~~ Y. Ohnuki, Unitary Representation of the Poinca\'e Group and Relativistic

Wave Equations (World Scientific, Singapore 1988).

\noindent
3.~~ M. Flato, D. Sternheimer and C. Fronsdal, Commun. Math. Phys.

 90(1983)563.

\noindent
4.~~ T. Itoh and K. Odaka, Fortschr. Phys. 39(1991)557.

\noindent
5.~~ G.W. Mackey, Induced Representaions of Groups and Quantum Mechanics

 (Benjamin, New York 1969).

\noindent
6.~~ C.J. Isham, in Relativisty, Group and Topology II (ed. B.S. de Witt and

 R. Stora, North-Holland, Amsterdam 1984), and references cited therein.

\noindent
7.~~ P.A.M. Dirac, Lectures on Quantum Mechanics (Yeshiva, New York 1964).

\noindent
8.~~ N.P. Landsman and N. Linden, Nucl. Phys. 34(1991)121,

Y. Ohnuki and S. Kitakado, J. Math. Phys. 34(1993)2827.

\noindent
9.~~ K. Odaka, in preparation.

\noindent
10. \hskip 1truemm M.S. Marinov and M.V. Terentyev,  Fortschr. Phys.
27(1979)511, and

 references cited therein.

\noindent
11. \hskip 1truemm D. McMullan and I. Tsutsui, PLY-MS-93-04(1993)
(To be published in Phy.

 Lett.).
\bye